\documentclass[conference]{IEEEtran}
\IEEEoverridecommandlockouts
\usepackage[numbers,sort&compress]{natbib}
\usepackage{amsmath,amssymb,amsfonts}
\usepackage{algorithmic}
\usepackage{graphicx}
\usepackage{textcomp}
\usepackage{xcolor}
\usepackage[normalem]{ulem}
\usepackage{orcidlink}
\usepackage{cleveref}
\usepackage{balance}
\usepackage{csquotes}
\usepackage{xspace}
\usepackage{booktabs}
\usepackage{siunitx}
\usepackage[most]{tcolorbox}

\makeatletter
\newcommand\Cshadowbox{\VerbBox\@Cshadowbox}
\def\@Cshadowbox#1{%
  \setbox\@fancybox\hbox{\fbox{#1}}%
  \leavevmode\vbox{%
    \offinterlineskip
    \dimen@=\shadowsize
    \advance\dimen@ .5\fboxrule
    \hbox{\copy\@fancybox\kern.5\fboxrule\lower\shadowsize\hbox{%
      \color{ShadowColor}\vrule \@height\ht\@fancybox \@depth\dp\@fancybox \@width\dimen@}}%
    \vskip\dimexpr-\dimen@+0.5\fboxrule\relax
    \moveright\shadowsize\vbox{%
      \color{ShadowColor}\hrule \@width\wd\@fancybox \@height\dimen@}}}
\makeatother
\colorlet{ShadowColor}{gray}

\def\BibTeX{{\rm B\kern-.05em{\sc i\kern-.025em b}\kern-.08em
    T\kern-.1667em\lower.7ex\hbox{E}\kern-.125emX}}

\def\codebased{code-based\xspace}

\makeatletter
\newcommand{\classifier}[1][]{%
  \def\flake@pre{}%
  \def\flake@suf{}%
  \in@{f}{#1}\ifin@\def\flake@pre{flakiness\space}\fi
  \in@{s}{#1}\ifin@\def\flake@suf{s}\fi
  \flake@pre detector\flake@suf\xspace}
\makeatother
\def\classification{detection\xspace}
\def\cidoft{\texttt{C-IDoFT}\xspace}
\def\idoft{\texttt{IDoFT}\xspace}
\def\flakeflagger{\texttt{FlakeFlagger}\xspace}
\def\flakyq{\texttt{FlakyQ}\xspace}
\def\flakify{\texttt{Flakify}\xspace}
\def\flakebench{\texttt{FlakeBench}\xspace}
\def\flakyxbert{\texttt{FlakyXBert}\xspace}
\def\flakifyplus{\texttt{Flakify+}\xspace}
\def\flakylensplus{\texttt{FlakyLens+}\xspace}
\def\flakylens{\texttt{FlakyLens}\xspace}
\def\flakeci{\texttt{FlakeCI}\xspace}
\def\flakycat{\texttt{FlakyCat}\xspace}
\def\idflakies{\texttt{iDFlakies}\xspace}
\def\msrflakiness{\texttt{msr4flakiness}\xspace}
\def\deflaker{\texttt{DeFlaker}\xspace}

\DeclareRobustCommand{\del}[1]{}

\begin{document}

\author{%
\IEEEauthorblockN{%
Ömer Oktay Gültekin~\orcidlink{0009-0004-0901-1360}\textsuperscript{1},
Alexander Berndt~\orcidlink{0009-0009-5248-6405}\textsuperscript{1,3},
Jonathan Bell~\orcidlink{0000-0002-1187-9298}\textsuperscript{2},
Thomas Bach~\orcidlink{0000-0002-9993-2814}\textsuperscript{3}, and
Sebastian Baltes~\orcidlink{0000-0002-2442-7522}\textsuperscript{1}%
}
\IEEEauthorblockA{%
\textsuperscript{1}Heidelberg University, Germany \quad
\textsuperscript{2}Northeastern University, USA \quad
\textsuperscript{3}SAP, Germany\\[2pt]
oemer.gueltekin@stud.uni-heidelberg.de, alexander.berndt@uni-heidelberg.de, j.bell@northeastern.edu,\\
thomas.bach03@sap.com, sebastian.baltes@uni-heidelberg.de%
}
}

\title{How Far Are We from Detecting Flaky Tests? On~the~Limits of Code-Based Detection}

\maketitle
\bstctlcite{IEEEtran:BSTcontrol}
\begin{abstract}
Flaky tests pass and fail on the same code version, weakening the signal of test results and disrupting continuous integration (CI) pipelines. Code-based \classifier[fs] report strong benchmark results, yet their use in practice remains limited. We argue that the field is studying the wrong problem: Flakiness is not a static property of test code, which often lacks the information needed to decide whether a test is flaky.
Analyzing three \codebased \classifier[s] operating on test code, we found that widely used benchmarks contain shortcuts that inflate reported F1 scores and that evaluation protocols overstate generalizability. To control for these shortcuts, we curated two datasets. %
The first, \cidoft{} (\num{54468} unit tests from 57 GitHub projects), keeps a developer-confirmed subset of \idoft's flaky tests and rebuilds only the non-flaky class from repeated executions instead of fixed versions of flaky tests. \cidoft is a controlled counterfactual, not a benchmark for reuse. Our CodeBERT reimplementations of two published \classifier[s] scored far above its constant baselines under the published cross-validation protocol but no better than them once projects were separated. The high scores rested on the labeling shortcut and the evaluation protocol, not on the test code. On \flakebench, a benchmark restricted to flakiness types typically recognizable from test code, and the same project-disjoint protocol, the models identified nearly all flaky tests.
The second dataset, mined from CI logs, contains 86 flaky end-to-end tests that passed and failed on the same commit.
The test code and CI log yielded a cause for 42\% of them; the other 58\% required further execution evidence. Rather than abandoning flakiness prediction, we reframe it around whether an observed \emph{failure} is flaky and how likely a \emph{test} is to fail given its execution environment. Our datasets and CI-mining method support this direction.
\end{abstract}

\section{Introduction}
Flaky tests impede continuous integration (CI). Flaky tests pass and fail on the same version of the code, so their failures no longer reliably signal a software defect~\cite{DBLP:conf/icst/LeinenEPSSJ24}. Thus, flaky tests waste developer time spent debugging spurious failures and erode trust in test results~\cite{DBLP:conf/sigsoft/EckPCB19}. The common mitigation is to repeatedly re-execute failing tests in the CI pipeline to separate flaky failures from real ones. Companies such as Google rely on this to keep flaky failures from blocking their pipelines~\cite{DBLP:conf/ftw/HoangB24}, and Meta reports that flakiness reduces test effectiveness and increases the cost of testing~\cite{DBLP:conf/icst/AlshahwanHM23}. However, repeatedly executing flaky tests is expensive at scale, and at Google flaky tests even undermine rerun-based test selection heuristics~\cite{DBLP:conf/icse/MemonGNDNSM17}. For example, practitioners at SAP report that \num{500} hours of compute per day go to rerun-based handling of flaky failures during SAP~HANA pre-submit testing on the main branch alone~\cite{DBLP:conf/esem/BerndtBB24}.

To avoid the cost of repeatedly executing flaky tests, a large body of work tries to detect flaky tests statically, without reruns, using signals such as the test execution history~\cite{DBLP:conf/ftw/HoangB24,DBLP:conf/icse/KowalczykNGSLM20,DBLP:conf/icse/HerzigN15}, code metrics~\cite{DBLP:conf/sigsoft/PontilloPF21}, or combinations of metrics and the test code~\cite{DBLP:journals/infsof/CaiLLLXW26}.
The most actively studied approach is \codebased \classification, where a model is trained on a test's source code to predict whether it is flaky, ranging from \citeauthor{DBLP:conf/msr/0001MDdTB20}'s lexical approach and follow-up vocabulary studies~\cite{DBLP:conf/msr/0001MDdTB20,DBLP:conf/msr/HabenHPCT21,DBLP:conf/esem/BerndtNB23,DBLP:conf/ftw/MedeirosM26} through feature-based and fine-tuned code models~\cite{DBLP:conf/icse/AlshammariMH021,DBLP:journals/tse/FatimaGB23,DBLP:journals/pacmpl/Rahman0S25} to recent methods based on Large Language Models (LLMs)~\cite{DBLP:conf/icst/MoreB25,DBLP:journals/corr/abs-2602-05465}, the strongest of which report F1 scores above 0.90 on standard benchmarks.
Yet this success has not reached practice, and industrial adoption remains limited~\cite{DBLP:journals/corr/abs-2602-05465,DBLP:conf/kbse/HabenHMHPCT24}. For example, in a previous industrial study, a \codebased \classifier was evaluated but never deployed because it did not generalize to real-world data~\cite{DBLP:conf/esem/BerndtNB23}.
Recent work has begun to explain why the benchmark numbers are fragile: Some are inflated by experimental flaws~\cite{DBLP:journals/pacmpl/Rahman0S25,DBLP:journals/corr/abs-2602-05465}, and the test code alone may not contain the information needed to separate flaky from non-flaky tests~\cite{DBLP:journals/corr/abs-2602-05465}.

We argue that these fragilities are symptoms of studying the wrong problem, not isolated defects to be patched.
Flakiness is not a static property of a test's source code: It arises from how the test interacts with the system under test and its execution environment at run time.
Asking \enquote{is this test flaky?} from the code alone is therefore ill-posed for many kinds of flakiness, no matter how the \classifier is built.
\citeauthor{DBLP:journals/pacmpl/Rahman0S25} traced one \classifier's inflated score to a data-leakage flaw in its cross-validation~\cite{DBLP:journals/pacmpl/Rahman0S25}.
We additionally found that the same data leakage affected that \classifier's published leave-one-project-out evaluation.
In this work, we broaden the critique from a single implementation defect to the task framing itself, and we reproduce published approaches under a single, rigorous protocol, project-disjoint wherever the approach permits it, showing that the failure of \codebased detection persists across benchmarks and models.

We proceed in two steps.
First, we reproduce three \codebased \classifier[s] published at major software engineering venues~\cite{DBLP:journals/tse/FatimaGB23,DBLP:conf/icst/RahmanBMS24,DBLP:conf/icst/MoreB25} and audit their artifacts and benchmarks.
We find that commonly used benchmarks contain shortcuts that inflate the reported F1 scores and that the evaluation protocols overstate generalizability.
To control for these shortcuts, we build \cidoft, a dataset of \num{54468} unit tests from 57 projects, with flaky labels from developer-confirmed tests in the International Dataset of Flaky Tests (\idoft) and non-flaky labels from repeated execution rather than from fixed versions of flaky tests (the \emph{fix-commit} shortcut).
Where an earlier critique demonstrated the problem on a newly constructed dataset~\cite{DBLP:journals/pacmpl/Rahman0S25}, \cidoft keeps the developer-confirmed flaky tests those \classifier[s] were evaluated on and rebuilds only their non-flaky counterparts from reruns.
After the shortcut was removed and projects were separated, the \classifier[s] did no better than a majority-class baseline, a constant predictor that always outputs the more frequent class. However, we argue that this weak performance is caused by the ill-posed framing of the task, and view \cidoft as a control rather than a benchmark we recommend adopting.

Second, we analyze whether CI logs, in addition to the test code, can be used as a signal to identify flaky failures. We mine more than \num{450000} CI runs from open-source GitHub projects to build \flakeci, a set of flaky end-to-end (E2E) tests that pass and fail on the same commit.
Based on this dataset, we could attribute a cause to 42\% of these tests from the test code and CI log together. For the remaining tests, diagnosis required execution evidence that neither source provided.

To summarize, these results relocate the problem: They move flaky test detection from the test's code to its execution and ask whether an observed \emph{failure} is flaky or how likely a \emph{test} is to fail.
Once the benchmark shortcuts are removed and projects are separated, code-based detectors are, on developer-confirmed flakiness with rerun-confirmed non-flaky labels, no closer to detecting flaky tests than a majority-class baseline, and the progress reported in prior studies largely rested on those shortcuts and the cross-validation protocol. We argue that this failure-level formulation is where the field can make genuine progress.

The contributions of this work are as follows:
\begin{enumerate}
    \item A critical analysis showing that the apparent success of \codebased flaky test \classification is largely an artifact of how the task and its benchmarks are framed, demonstrated by reproducing three published \classifier[s].
    \item \cidoft, a controlled counterfactual that rebuilds the non-flaky class of the \idoft benchmark used in prior studies, removing the fix-commit shortcut.
    \item \flakeci, a CI-log-mined corpus of confirmed flaky E2E failures, with evidence that for most of its tests, a cause could not be attributed from the test code and the CI log alone.
\end{enumerate}

The remainder of this paper is structured as follows: \Cref{sec:background} introduces background on flaky test \classification and the studies we reproduce. \Cref{sec:methodology} introduces our research questions and methodology. \Cref{sec:results} presents our empirical results. We discuss the results in \Cref{sec:discussion} and threats to their validity in \Cref{sec:threats}. \Cref{sec:related} discusses related work. \Cref{sec:conclusion} concludes this work. Our replication package is available~\cite{zenodo}.

\section{Background}
\label{sec:background}

\subsection{Flaky Test Detection}
Previous work, surveyed systematically by \citeauthor{DBLP:journals/tosem/ParryKHM22}~\cite{DBLP:journals/tosem/ParryKHM22}, proposed approaches that detect flaky tests statically, i.e., without repeated executions. These approaches can be grouped by the data they use for \classification: the test execution history~\cite{DBLP:conf/ftw/HoangB24,DBLP:conf/icse/KowalczykNGSLM20},
metrics of the test such as lines of code or execution time~\cite{DBLP:conf/icse/AlshammariMH021,DBLP:conf/sigsoft/PontilloPF21}, or the test's source code itself, from vocabulary-based models~\cite{DBLP:conf/msr/0001MDdTB20,DBLP:journals/access/VerdecchiaCMB21} and their replications~\cite{DBLP:conf/msr/HabenHPCT21,DBLP:conf/iwpc/CamaraSEV21,DBLP:conf/esem/BerndtNB23} to CodeBERT-based \classifier[s]~\cite{DBLP:conf/icst/RahmanBMS24,DBLP:conf/icst/MoreB25,DBLP:journals/infsof/CaiLLLXW26}.
In this work, we focus on the latter group, the \codebased approaches operating on the test's source code, from the vocabulary-based models to the CodeBERT-based \classifier[s], which have recently received increasing attention due to their promising results across various benchmark datasets.

The intuition behind \codebased flaky test \classification is that flaky tests exhibit syntactical patterns that distinguish them from non-flaky tests~\cite{DBLP:conf/msr/0001MDdTB20}.
After \citeauthor{DBLP:conf/msr/0001MDdTB20} introduced the approach in \citeyear{DBLP:conf/msr/0001MDdTB20}~\cite{DBLP:conf/msr/0001MDdTB20}, replications revealed generalizability issues.
Performance degraded on newer revisions of a repository (up to a 21\% drop in MCC)~\cite{DBLP:conf/msr/HabenHPCT21} and on unseen projects~\cite{DBLP:conf/iwpc/CamaraSEV21}.
An industrial evaluation concluded that this lack of generalizability prevents production deployment~\cite{DBLP:conf/esem/BerndtNB23}.
Therefore, more recent work has proposed using LLMs for flaky test \classification~\cite{DBLP:journals/tse/FatimaGB23,DBLP:conf/icst/RahmanBMS24,DBLP:conf/icst/MoreB25}, with \flakify explicitly motivated by these generalizability issues. \citeauthor{DBLP:journals/tse/FatimaHB24} extended this line to predicting flaky-test fix categories with LLMs~\cite{DBLP:journals/tse/FatimaHB24}.

\subsection{Datasets}
\label{sec:datasets}

The datasets used to evaluate \codebased flaky test \classification share a common lineage but differ along two independent axes in how their flaky and non-flaky labels are obtained. On the flaky axis, tests are either (i) mined from community reports or curated tool runs, (ii) drawn from an expert-curated category set, (iii) detected by reruns, or (iv) mined from CI logs (the strategy behind our \flakeci dataset, see \Cref{sec:methodology}). On the non-flaky axis, a test is labeled non-flaky in one of two ways: (i) the test is a developer-fixed later version of a reported flaky test (\emph{fix-commit}), or (ii) it survives many repeated executions (\emph{rerun-confirmed}). Each dataset pairs one flaky source with one non-flaky source. We organize the datasets below by that pairing and describe them in chronological order.

\textbf{msr4flakiness.} \citeauthor{DBLP:conf/msr/0001MDdTB20} constructed the first \codebased flakiness dataset by running every test case of 24 projects from the \deflaker{} benchmark~\cite{DBLP:conf/icse/BellLHEYM18} 100 times, a total of \num{64000} tests and roughly 6.4 million test executions. Tests that did not change their outcome across reruns were labeled non-flaky~\cite{DBLP:conf/msr/0001MDdTB20}.
This dataset uses rerun-confirmed non-flaky labels with a 100-run budget. Its flaky labels are inherited from the \deflaker{} dataset.

\textbf{FlakeFlagger.} \flakeflagger selected 24 projects previously studied in flaky test research, 17 of which overlap with those of \msrflakiness~\cite{DBLP:conf/icse/AlshammariMH021}.
To obtain rerun-based labels with a deeper budget, the authors re-executed each project's test suite \num{10000} times.
They found that even \num{10000} reruns did not detect all flaky tests, and that shallower budgets miss many tests that fail only rarely.
As reported by \citeauthor{DBLP:journals/tse/FatimaGB23}, the resulting dataset contains 802 flaky and \num{20859} non-flaky tests across 23 projects, totaling \num{21661} tests~\cite{DBLP:journals/tse/FatimaGB23}.
In addition to rerun-based labels, \flakeflagger provides features such as test smells, coverage, and execution time.
Because both its flaky and non-flaky labels come from the same deep rerun campaign rather than from developer fixes or an externally sourced flaky list, \flakeflagger is the most realistic of the public datasets and the one we do \emph{not} modify in this work.

\textbf{IDoFT.} The International Dataset of Flaky Tests (\idoft) is an open-source GitHub repository tracking flaky tests in open-source Java and Python projects~\cite{InternationalDatasetofFlakyTests,LamETAL19iDFlakies}.
Its flaky tests stem from research detection tools in the lineage of \idflakies, which detects flaky tests by rerunning a project's tests under reordered execution orders, and from community submissions~\cite{LamETAL19iDFlakies}.
\idoft itself lists only flaky tests, together with each test's fix status. The classification dataset derived from it, introduced by \flakify~\cite{DBLP:journals/tse/FatimaGB23}, adds \emph{fix-commit} non-flaky labels. A non-flaky test is the fixed version of a flaky test, obtained from the pull request linked in \idoft, where the fix changed the test code itself.
The variant used to evaluate \flakify and its successors consists of \num{3813} tests, of which \num{3195} are flaky and 618 are non-flaky~\cite{DBLP:conf/icst/RahmanBMS24}.\footnote{\label{fn:idoft-splits}\flakify's original evaluation used \num{3862} tests (\num{3268} flaky, 594 non-flaky), consistent between its paper and its replication package. The \num{3813}-test variant is the one shipped in \flakyq's replication package, which we and the reproduced studies use. Its label column contains \num{3226} flaky and 587 non-flaky tests rather than the reported \num{3195} and 618.}
As we discuss in \Cref{sec:methodology}, this fix-commit construction yields two artifacts: a non-flaky class far smaller than the flaky class, and fixed/unfixed test pairs that are nearly identical in their code.

\textbf{FlakeBench.} \flakebench~\cite{DBLP:journals/pacmpl/Rahman0S25} was built to give \flakycat's flaky tests a realistic non-flaky population.
\flakycat{} is a manually curated dataset of categorized flaky tests~\cite{akli2023flakycat} whose categories derive from root causes defined in prior studies~\cite{DBLP:conf/sigsoft/LuoHEM14,DBLP:conf/sigsoft/EckPCB19}.
To build the non-flaky population, \citeauthor{DBLP:journals/pacmpl/Rahman0S25} started with projects evaluated by \flakycat{} and extracted \num{179403} candidate tests. After removing all tests that share a test class with a known flaky test, they ran each remaining candidate 100 times to confirm it was non-flaky.
The resulting dataset contains \num{8574} tests, of which 280 are flaky (\num{3.3}\% across the five flakiness categories \flakebench retains from \flakycat\footnote{\citeauthor{DBLP:journals/pacmpl/Rahman0S25} state \num{3.4}\%; 280 of \num{8574} tests is \num{3.3}\%, whereas \num{3.4}\% corresponds to the ratio of flaky to non-flaky tests (280/\num{8294}).}) and \num{8294} are non-flaky, spanning 98 projects.\footnote{\citeauthor{DBLP:journals/pacmpl/Rahman0S25} report 97 projects; the shipped dataset spans 98 projects, one of which contributes only non-flaky tests.}
\flakebench thus pairs report/category-mined flaky labels with rerun-confirmed non-flaky labels.

The contrast between these strategies motivates our own \cidoft: Rather than inheriting \idoft's fix-commit non-flaky labels, we extend \idoft by deriving non-flaky tests from repeated executions of the modules that contain flaky tests, similar to \flakebench's rerun-confirmed non-flaky construction~\cite{DBLP:journals/pacmpl/Rahman0S25} (\Cref{sec:methodology}).
\flakebench and \cidoft share the rerun-confirmed non-flaky rule, but differ in what they hold fixed, and the difference determines what each can establish.
\flakebench substitutes the flaky population (drawing from \flakycat{} and discarding the Network, Randomness, and Resource Leak categories), the projects, and the non-flaky construction together, producing a separate benchmark.
A separate benchmark confounds the question we set out to answer: Lower performance could mean its tests are harder to classify or differently distributed.
To attribute a change in performance to the labeling rather than to the dataset, the flaky tests must be held fixed while only the non-flaky construction varies.
\cidoft is built for exactly this purpose: It uses the same \idoft benchmark that the \classifier[s] we reproduce were evaluated on, restricts its flaky tests to developer-confirmed reports, and changes only how the non-flaky class is constructed.

\subsection{Reproduced Detectors}
We reproduce three existing approaches that use LLMs for \codebased flaky test \classification~\cite{DBLP:journals/tse/FatimaGB23,DBLP:conf/icst/RahmanBMS24,DBLP:conf/icst/MoreB25}. There are other studies on \codebased flaky test \classification, such as vocabulary-based token models~\cite{DBLP:journals/access/VerdecchiaCMB21}, combinations of expert features with CodeBERT embeddings~\cite{DBLP:journals/infsof/CaiLLLXW26}, and classification from statically extracted features~\cite{DBLP:conf/ftw/MedeirosM26}. We focused on \flakify, \flakyq, and \flakyxbert because they are LLM-based, were published at major SE venues (TSE and ICST), and provide replication packages.

\textbf{Flakify} uses a fine-tuned version of \emph{CodeBERT}, a pre-trained language model with 125 million parameters~\cite{DBLP:journals/tse/FatimaGB23,DBLP:conf/emnlp/FengGTDFGS0LJZ20}.
Under cross-validation, \flakify reports an almost perfect F1 score of 0.98 on \idoft but only 0.79 on the rerun-labeled \flakeflagger dataset; under per-project validation, the gap persists (0.89 on \idoft versus 0.73 on \flakeflagger).
In a follow-up study, \citeauthor{DBLP:journals/pacmpl/Rahman0S25} found that \flakify's implementation suffered from data leakage in its cross-validation setup, where a single CodeBERT instance was reused across folds rather than reloaded, inflating its reported F1 score~\cite{DBLP:journals/pacmpl/Rahman0S25}.
We found that the same data leakage also affected \flakify's published leave-one-project-out evaluation: Projects were evaluated sequentially, and the model was not reinitialized between held-out projects, so later evaluations could reuse weights already updated on earlier runs. The results without this leakage remained unknown. We fill this gap by replicating \flakify with pipelines that reload the model for each fold or held-out project.

\textbf{FlakyQ} is a quantized version of \flakify that improves inference efficiency.
\flakyq achieved an F1 score of up to 0.94 on \idoft while reducing prediction time and RAM usage by 25\% and 48\%~\cite{DBLP:conf/icst/RahmanBMS24}.
Its evaluation uses the \idoft variant (\num{3813} tests) and the \flakycat{} categorized flaky tests, not the rerun-labeled \flakeflagger set.
As \flakyq reuses \flakify's training and evaluation pipeline, adding quantization for inference, it suffers from the same data leakage.

\textbf{FlakyXBert} uses few-shot learning (FSL) with a Siamese network to train a CodeBERT-based \classifier[f]~\cite{DBLP:conf/icst/MoreB25}.
\citeauthor{DBLP:conf/icst/MoreB25} concluded that FSL yields a competitive F1 score compared with fine-tuning, while requiring less data and fewer computational resources, and thus may be better suited for settings where the model must adapt rapidly to changes in the code base without extensive retraining.
\flakyxbert is evaluated on the \idoft variant and the \flakycat{} dataset.

Of the three reproduced approaches, only \flakify reports results on the rerun-labeled \flakeflagger dataset, and that is precisely the dataset on which it scores lowest (0.79 versus 0.98 on \idoft).
\flakylens~\cite{DBLP:journals/pacmpl/Rahman0S25}\footnote{Our replication package additionally includes \flakylensplus{} evaluations (\Cref{sec:methodology}) on \cidoft, \flakeflagger, and \flakebench, provided for completeness, together with four-class category classification runs on \cidoft, which this paper does not analyze because they address a different task: predicting a test's flakiness category rather than the binary \classification our research questions study; \flakylens{} is not among the three \classifier[s] we reproduce.}, a focal-loss CodeBERT \classifier[f], is introduced together with \flakebench and evaluated on it; neither \flakyq, \flakyxbert, nor \flakylens{} reports results on \flakeflagger's rerun-based labels.
The benchmarks the field continued to use are those whose non-flaky labels do not come from deep reruns (the \idoft variant's fix-commit labels or \flakebench's 100-run confirmation), while the dataset whose non-flaky labels come purely from deep reruns is the one on which the strongest reported \classifier already struggled. We add rerun-confirmed non-flaky tests to the \idoft benchmark to see whether this influences the performance reported in prior studies.

\subsection{Flaws in Prior Work and the Gap We Address}
\label{sec:gap}
Prior work has identified a consistent set of flaws in \codebased flaky test \classification, but reports each as an isolated defect.
The trained models do not generalize, degrading on newer project revisions~\cite{DBLP:conf/msr/HabenHPCT21}, on unseen projects~\cite{DBLP:conf/iwpc/CamaraSEV21}, and on industrial data, where one \classifier was evaluated but never deployed~\cite{DBLP:conf/esem/BerndtNB23}. Some reported gains are also experimental artifacts, such as the cross-validation data leakage \citeauthor{DBLP:journals/pacmpl/Rahman0S25} traced through a prior \classifier~\cite{DBLP:journals/pacmpl/Rahman0S25}.
A deeper problem is that the test code may not carry the signal at all: The widely used \idoft benchmark pairs each flaky test with a near-identical developer-fixed version, so a model can separate the classes from superficial similarity rather than from flakiness. Even human experts often do not consider themselves able to label a test from its code alone~\cite{DBLP:journals/corr/abs-2602-05465}.

Each flaw has been treated as something to patch, a leak to fix or a benchmark to clean, leaving open whether the task itself is well-posed. Answering that requires a controlled counterfactual without the shortcut, which is a central contribution of our study.
We construct \cidoft for this purpose (\Cref{sec:datasets}), removing the fix-commit shortcut while restricting the flaky tests to a developer-confirmed subset of the same benchmark.
If \codebased \classifier[s] still fail on \cidoft, the limitation lies in the task rather than in flaws of a particular dataset. \cidoft thus serves as a control, not as a benchmark we recommend adopting.
We then construct \flakeci, a dataset of flaky E2E failures mined from CI logs, to examine whether the evidence for diagnosing a flaky failure lies in the test code, in the CI log, or beyond both.
Together these datasets let us reproduce three published LLM-based \classifier[s] under a single rigorous protocol, project-disjoint wherever the design permits it, and ask whether the test code contains the discriminating signal at all, rather than only whether their reported F1 scores can be reproduced.

\section{Study Design}
\label{sec:methodology}
In the following, we introduce our research questions and the methodology to answer them.

\begin{quote}
    \textbf{RQ1}: \emph{To what degree can we replicate the performance of existing \codebased flaky test \classifier[s]?}
\end{quote}

\textbf{Motivation.} Despite recurring promising results reported by studies on \codebased flaky test \classification, recent work cast doubts on the generalizability of the trained models and their applicability in practice~\cite{DBLP:journals/corr/abs-2602-05465,DBLP:journals/pacmpl/Rahman0S25}.
While the original vocabulary-based approach has been replicated on newer revisions, unseen projects, and industrial data (\Cref{sec:background}), there is no evidence on whether the identified generalizability issues also apply to \codebased \classifier[s] that use LLMs.

\textbf{Approach.} We adhere to ACM's definition for artifact review and aim to first reproduce the results of the given studies using the author-supplied artifacts in the replication packages of the respective studies~\cite{ACMBadging2020}. We focus our reproduction on the experiments related to flakiness \classification.
For a sound comparison, we use the same performance metrics as the original papers, namely the weighted F1 score and the flaky-class F1 score. Given the precision and recall for a class $c$, the F1 score for $c$ is the harmonic mean $F1_c = \frac{2\cdot\text{Precision}_c\cdot\text{Recall}_c}{\text{Precision}_c \mathbin{+} \text{Recall}_c}$. %
The flaky-class F1 score $F1_f$ is $F1_c$ evaluated for the flaky class. The weighted F1 score $F1_w$ is the weighted average over a set of classes $C$, where each class $c \in C$ has support $n_c$ and $N=\sum_{c\in C} n_c$, such that $F1_w=\sum_{c\in C} \frac{n_c}{N} \cdot F1_c$. $F1_w$ is dominated by whichever class holds the majority: the flaky class on the \idoft variant used in \textbf{RQ1}, whose non-flaky tests are fix-commit versions and where \num{3195} of \num{3813} tests are flaky, and the non-flaky class on the rerun-labeled datasets, where only 2--4\% of tests are flaky. On the latter, $F1_w$ can approach $1$ even if the \classifier detects almost no flaky tests. We therefore report the flaky-class metrics as the more informative measures of detection and compare every result against a majority-class baseline, which always predicts flaky or always predicts non-flaky, depending on the dataset's majority class. On the flaky-class metrics, we additionally report the always-flaky predictor, whose flaky-class F1 score is the maximum any input-agnostic predictor can attain (\(F1_f = 2p/(1+p)\) at recall 1, for flaky share \(p\)). Cross-validation results are reported as the fold mean, with $\pm$ denoting the fold-to-fold standard deviation. In the project-disjoint setting, no project contributes tests to both the training and the test set of a given fold, so the reported performance reflects generalization to unseen projects.
After reproducing the results using the original pipelines, we replicated the proposed approaches in the project-disjoint setting defined above.
This is similar to the setup of \citeauthor{DBLP:conf/iwpc/CamaraSEV21}~\cite{DBLP:conf/iwpc/CamaraSEV21}, who found that this setting leads to a performance drop for the lexical \codebased models.

In addition to the quantitative results from the evaluation pipelines, we performed a qualitative analysis of the datasets to understand the characteristics that LLMs can use to distinguish flaky from non-flaky tests given these benchmarks.

\begin{quote}
    \textbf{RQ2}: \emph{What is the performance of CodeBERT-based \classifier[s] on a controlled-counterfactual dataset that removes the fix-commit labeling shortcut?}
\end{quote}

\textbf{Motivation.} Previous work pointed out that the \idoft benchmark, which \flakify~\cite{DBLP:journals/tse/FatimaGB23}, \flakyxbert~\cite{DBLP:conf/icst/MoreB25}, and \flakyq~\cite{DBLP:conf/icst/RahmanBMS24} used for evaluation, exhibits spurious features that can be exploited by a trained model to distinguish non-flaky and flaky tests~\cite{DBLP:journals/corr/abs-2602-05465}.
The authors traced this to the dataset's construction: Non-flaky samples are the fixed versions of previously flaky tests, so a model can separate the two classes from characteristics the fix introduces, such as the shared fix code itself, rather than from any signal of flakiness.
\flakebench responded to this problem with a new dataset whose flaky tests come from a different source and different projects. \cidoft instead keeps a developer-confirmed subset of \idoft's flaky tests and changes only the non-flaky labeling, so that a change in performance is attributable to the labeling and not to the different set of tests a new dataset would introduce (\Cref{sec:datasets}).

\textbf{Approach.} To mitigate the fix-commit shortcut, we first curated a new dataset from $n=500$ repeated test executions across the projects in the \idoft benchmark, a common approach for labeling tests in flakiness datasets~\cite{DBLP:conf/icse/GruberRPSM024,DBLP:conf/icse/AlshammariMH021}. The run budget of $n=500$ was chosen to balance statistical confidence with the computational cost of module-level execution. Given a test with an independent per-run failure probability $p$, the probability of observing at least one failure in $n$ runs is $1-(1-p)^n$. With $n = 500$, a test with a failure probability of 1\% has a 99.3\% chance of failing at least once~\cite{DBLP:journals/corr/abs-2601-08998}.
After repeatedly executing the tests and labeling them, we fine-tuned our two CodeBERT reimplementations, \flakifyplus{} (class-weighted cross-entropy, following \flakify's training setup) and \flakylensplus{} (focal loss, following \flakylens), on \cidoft and evaluated their performance.
For evaluation, we report the flaky-class precision, recall, and F1 score as our primary metrics, and report the weighted F1 score only for comparison with the published results, as motivated above.
We additionally compare against a majority-class (always-non-flaky) baseline.
As in \textbf{RQ1}, we evaluated in a project-disjoint setting, holding out entire projects from training.
\Cref{tab:dataset-suite} summarizes the sizes of the datasets used to answer \textbf{RQ1} and \textbf{RQ2}.

\begin{table}
  \centering
  \caption{The three datasets used to answer \textbf{RQ1} and \textbf{RQ2}, with their project and flaky/non-flaky test counts. For \flakeflagger, we list the \num{20858} non-flaky tests that enter our project-disjoint evaluation; the original paper reports \num{20859} before removing one test without a project label~\cite{DBLP:journals/tse/FatimaGB23}. The \idoft row lists the split reported by \flakyq (see footnote~\ref{fn:idoft-splits}).}
  \label{tab:dataset-suite}
  \small
  \begin{tabular}{@{}p{0.15\textwidth}rrr@{}}%
    \toprule
    \textbf{Dataset} & \textbf{Projects} & \textbf{Flaky} & \textbf{Non-flaky}\\
    \midrule
    \flakeflagger{} & 23 & 802 & \num{20858} \\
    \cidoft{} & 57 & \num{1319} & \num{53149} \\
    \idoft{} & 126 & \num{3195} & \num{618} \\
    \bottomrule
  \end{tabular}
\end{table}

\begin{quote}
    \textbf{RQ3}: \emph{To what degree can we diagnose flaky failures in E2E tests given the test code and CI logs?}
\end{quote}

\textbf{Motivation.} Flakiness is especially costly in slow and expensive E2E test suites, where extensive reruns are impractical, and failures are hard to diagnose~\cite{DBLP:conf/kbse/NgoNN22,DBLP:conf/icse/BerndtBB24}. In JavaScript, the language of the E2E suites we study, flaky tests are dominated by concurrency, async-wait, OS, and network causes~\cite{DBLP:conf/icsm/HashemiTR22}.
Yet the projects that make up existing flaky-test benchmarks are, by construction, those that can be built and executed quickly enough to be rerun at scale, often with build times of only a few minutes, which biases the field toward rather simple unit tests and away from the complex systems where flakiness is most problematic and expensive~\cite{DBLP:conf/icst/AlshahwanHM23}.
Project selection thus matters as much as test selection.
If flakiness in these E2E settings is driven by factors not observable in the test source code, then \codebased \classification is the wrong approach precisely where it matters most.
This concern is consistent with recent evidence that even human experts often do not consider flakiness identifiable from the test code alone, depending on the type of flakiness~\cite{DBLP:journals/corr/abs-2602-05465}.

The flakiness of E2E tests has different causes from those of the small unit tests of \textbf{RQ1} and \textbf{RQ2}, so simply reporting that a \classifier scores poorly on them would prove little, since the drop could reflect a harder and different domain rather than the task framing, the same confounding factor we avoid in \textbf{RQ2}.
We therefore do not evaluate \codebased \classification performance on these tests.
Instead, we manually assess whether the evidence explaining a flaky E2E failure is present in the test code and CI logs.

\textbf{Approach.} In contrast to prior work, which identified flaky E2E tests in open-source projects based on flakiness-related keywords in commits and closed issues~\cite{DBLP:conf/icse/RomanoSGYW21}, or detected build-level flakiness from restarted CI builds whose outcome changed on the same code~\cite{DBLP:conf/msr/DurieuxGHA20}, we constructed \flakeci by mining the GitHub Actions CI logs of open-source Cypress and Playwright projects, identifying E2E tests that both pass and fail on the same commit within the same workflow and job context, so that outcome differences produced by unrelated configuration changes, such as different runner images or environment variables, are excluded. One author assigned each of the 86 confirmed flaky tests a primary cause category from the test title, test code, and the matched failing and passing log evidence. The annotating author discussed the category set and the annotation procedure with a second author and assigned a test to the \emph{Unknown} category whenever the evidence did not support a cause without speculation. We then quantified the share of flaky failures whose diagnosis depends on execution evidence beyond the test code and CI logs.

\section{Results}
\label{sec:results}
All experiments were executed on an Apple Mac Studio with an Apple M3 Ultra chip (32-core CPU, 80-core GPU) and 512 GB of unified memory.

\subsection{RQ1: Replication}
With the first research question, we investigated whether the previous results of existing \codebased \classifier[s] could be validated in a more rigorous evaluation. As described in \Cref{sec:methodology}, we started by reproducing the existing studies given the author-supplied replication packages. We report the results of reproducing \flakify and \flakyq on the \idoft benchmark in \Cref{tab:rq1-idoft}. Run as shipped, the replication packages yielded results similar to the values reported in the paper, with differences of at most 0.02. After we fixed the data leakage, the weighted F1 score dropped from 0.95 to 0.88 for both \flakify and \flakyq. The per-fold flaky-class F1 score showed the same drop (\flakify{} 0.975 $\pm$ 0.014 to 0.919 $\pm$ 0.026; \flakyq{} 0.969 $\pm$ 0.019 to 0.925 $\pm$ 0.025) and remained high, as the flaky class is the majority on this dataset. Under project-disjoint evaluation, the weighted F1 score fell further to 0.80, matching the always-flaky baseline of 0.80 on the same held-out set. As mentioned in \Cref{sec:background}, in addition to \idoft, \flakify was evaluated on \flakeflagger. We report the results before and after fixing the data leakage in \Cref{tab:flakeflagger-lopo}. The \flakeflagger results repeated the \idoft pattern: The flaky-class F1 score collapsed after fixing the leakage (from 0.79 to 0.64) and again under project-disjoint evaluation (to 0.07). Under the same per-project aggregation, the always-flaky baseline reached 0.14, so the \classifier did not exceed it.

For \flakyxbert, we noticed a mismatch between the projects reported in the paper and those analyzed in the notebooks.
Therefore, we report the per-project results of \flakyxbert's reproduction in \Cref{tab:xbert-idoft}. Most of the F1 scores showed only minor deviations. However, the outputs shipped in four of the notebooks stem from another project's run, leading to an inflated performance estimate in all cases.\footnote{In all four cases, the notebook's code names the correct project, but its saved outputs, including the per-sample predictions, are identical to those of another project's notebook: \texttt{commons-lang} and \texttt{flink} match \texttt{spring-data-r2dbc}, \texttt{fastjson} matches \texttt{admiral}, and \texttt{graylog2-server} matches \texttt{mockserver}. Whether the wrong data was loaded at run time or outputs from another run were retained cannot be determined from the released artifact.} With the notebooks rerun on the correct data, the F1 score dropped from 0.94, which closely resembles the value reported in the paper, to 0.83.

Since \flakyxbert's authors did not include the results from all projects in their paper, restricting the aggregate to the projects reported there yielded an F1 score of 0.88, while the paper reported 0.95. After we notified them, the authors acknowledged this mistake in the paper and confirmed the F1 score reported in the notebooks.
Unlike \flakify and \flakyq, \flakyxbert cannot be replicated under our project-disjoint protocol: Its few-shot design requires labeled examples from the target project, so its per-project scores measure within-project classification and leave cross-project generalization untested.

Beyond the four notebooks with another project's outputs, our qualitative analysis revealed a preprocessing shortcut. The \flakify pipeline keeps only statements associated with targeted test smells, so tests without such smells collapse to empty method bodies. In \texttt{adyen-java-api-library}, 76 of 89 tests in the shipped CSV have such empty method bodies; 36 of them are empty because their rows are corrupted, not because of the preprocessing. We repaired these rows from the pipeline's source dataset to check whether the corruption explained \flakyxbert's low score on this project: The per-project F1 score rose from 0.31 to 0.94. The gain, however, rests on the shortcut, not on flakiness detection. After the repair, 32 of 35 non-flaky training examples have empty method bodies and none of the 36 flaky ones do, so separating empty from non-empty methods suffices.

\begin{tcolorbox}[enhanced jigsaw,sharp corners, drop fuzzy shadow=ShadowColor]
\noindent\textbf{Answer RQ1}:
The results of \flakify and \flakyq were reproducible within 0.02 with the author-supplied replication packages. After fixing the data leakage, the weighted F1 score dropped from 0.95 to 0.88 for both approaches, and under project-disjoint evaluation neither meaningfully exceeded the always-flaky baseline: 0.80 versus 0.80 in the weighted F1 score on the held-out \idoft projects, and 0.07 versus 0.14 in the per-project flaky-class F1 score on \flakeflagger.
For \flakyxbert, four notebooks ship outputs from another project's run, inflating the F1 score; after correction, its weighted F1 score dropped from 0.94 to 0.83, remaining above the always-flaky baseline of 0.66, unlike \flakify and \flakyq. Its few-shot design trains on labeled tests from the project it classifies, so 0.83 is not a cross-project result.
\end{tcolorbox}

\begin{table}
  \centering
  \caption{Average weighted F1 score in 10-fold cross-validation on the \idoft dataset before and after fixing the data leakage. Paper: the studies' reported values. Prefix and Postfix: our reproduction before and after this fix (model reloaded per fold). Disjoint: the project-disjoint evaluation. Baseline: the always-flaky weighted F1 score on the full \idoft{} (\num{3813} tests, labeled as shipped; see footnote~\ref{fn:idoft-splits}); on Disjoint's held-out \num{2105} tests, this baseline is 0.80.}
  \label{tab:rq1-idoft}
  \small
  \setlength{\tabcolsep}{2.8pt}%
  \begin{tabular}{lrrrrr}
    \toprule
    \textbf{Classifier} & \textbf{Paper} & \textbf{Prefix} & \textbf{Postfix} & \textbf{Disjoint} & \textbf{Baseline}\\
    \midrule
    \flakify & 0.94 & 0.95 $\pm$ 0.02 & 0.88 $\pm$ 0.03 & 0.80 & 0.78 \\
    \flakyq & 0.93 & 0.95 $\pm$ 0.03 & 0.88 $\pm$ 0.03 & 0.80 & 0.78 \\
    \bottomrule
  \end{tabular}
\end{table}

\begin{table}
\centering
\caption{Weighted F1 score in per-project evaluation of \flakyxbert on the \idoft dataset. Shipped: the values rendered in the replication package's notebooks, not reported in the paper. For projects marked with $\dagger$, the shipped notebooks contain outputs from another project's run rather than from the project's own data. Paper: the paper's reported values. Rerun: each project's notebook re-executed on its own data. Weighted Average: per-project values aggregated by support.}
\label{tab:xbert-idoft}
\footnotesize
\setlength{\tabcolsep}{3pt}
\begin{tabular}{lrrrr}
\toprule
\textbf{Project} & \textbf{Rerun} & \textbf{Shipped} & \textbf{Paper} & \textbf{Always flaky} \\ \midrule
\texttt{junit-quickcheck} & 0.92 & 0.94 & 0.94 & 0.36\\
\texttt{dubbo} & 0.89 & 0.89 & 0.89 & 0.88\\
\texttt{hadoop} & 0.95 & 0.95 & 0.95 & 0.97\\
\texttt{nifi} & 0.92 & 0.92 & 0.92 & 0.93\\
\texttt{admiral} & 0.91 & 0.91 & 0.91 & 0.95\\
\texttt{fastjson}$^\dagger$ & 0.55 & 0.91 & 0.91 & 0.43\\
\texttt{spring-data-r2dbc} & 1.00 & 1.00 & 1.00 & 0.38\\
\texttt{mockserver} & 1.00 & 1.00 & 1.00 & 0.67\\
\texttt{commons-lang}$^\dagger$ & 0.40 & 1.00 & -- & 0.45\\
\texttt{flink}$^\dagger$ & 0.71 & 1.00 & -- & 0.79\\
\texttt{graylog2-server}$^\dagger$ & 0.50 & 1.00 & -- & 0.46\\
\texttt{adyen-java-api-library} & 0.31 & 0.31 & 0.30 & 0.34\\
\midrule
\textbf{Weighted Average} & \textbf{0.83} & \textbf{0.94} & \textbf{0.95} & \textbf{0.66}\\ \bottomrule
\end{tabular}
\end{table}

\begin{table}
  \centering
  \caption{\flakify on \flakeflagger, reported as the averaged flaky-class F1 score from cross-validation, following the original paper. The project-disjoint rows and the always-flaky baseline are per-project means.}
  \label{tab:flakeflagger-lopo}
  \small
  \begin{tabular}{lr}
    \toprule
    \textbf{Evaluation} & \textbf{Flaky-class F1 score} \\
    \midrule
    Paper Average & 0.79 \\
    Corrected Average & 0.64 \\
    Project-disjoint, per-project mean & 0.07 \\
    Always flaky, per-project mean & 0.14\\
    \bottomrule
  \end{tabular}
\end{table}

\subsection{RQ2: A Controlled Counterfactual (\cidoft)}
\label{sec:rq2}
With the second research question, we aimed to find out whether our observations from \textbf{RQ1} hold on a controlled-counterfactual dataset that does not use fixed flaky tests as non-flaky samples.
We began with the \idoft dataset as used by \flakify. \flakify's authors used all \idoft flaky tests, including those whose flakiness
was reported but never confirmed by a developer.
We restricted the flaky tests in \cidoft to developer-confirmed tests from \idoft. We refer to a test as \emph{developer-confirmed} when the resolution status of its linked pull request is one of \emph{Accepted}, \emph{DeveloperFixed}, or \emph{InspiredAFix} (status values of \idoft's PR tracking), indicating that the reported flakiness was confirmed by a developer rather than merely reported or automatically detected. We first kept projects with a publicly accessible, non-archived, non-fork GitHub repository. We then required at least 10 developer-confirmed flaky tests per project, which left 69 candidate projects, of which 57 could be built and executed at the commits where the flakiness had been reported. From these 57 projects, we executed the modules containing developer-confirmed flaky tests 500 times. We labeled a test as non-flaky when all 500 executions yielded consistent results.

As model input, we used the full test code, filtered to at most 1024 CodeBERT tokens, avoiding smell-based preprocessing and the empty method bodies it induces. The 1024-token cap doubles \flakify's 512-token input limit while remaining compatible with CodeBERT's 512-position maximum, which our trainer handles by averaging two 512-token chunks; the cap retains 97.3\% of the extracted tests. We evaluated both variants with four project-disjoint folds, fine-tuning CodeBERT\footnote{Checkpoint \texttt{microsoft/codebert-base}, \url{https://huggingface.co/microsoft/codebert-base}.} for up to 20 epochs with early stopping (patience 5, batch size 4, learning rate $10^{-5}$, seed 42). The training folds drew their non-flaky class from a source deduplicated by template body (\num{42527} of \num{54468} rows, retaining all \num{1319} flaky tests); all reported metrics were computed on the full held-out test folds.

\Cref{tab:rq2} shows our project-disjoint results on \cidoft and \flakebench. On \cidoft, both CodeBERT variants collapsed: \flakifyplus reached a flaky-class F1 score of 0.035 (recall 0.021) and \flakylensplus 0.070 (recall 0.120), so both missed almost every flaky test. Neither meaningfully exceeded the constant baselines. \flakifyplus stayed below the always-flaky predictor's flaky-class F1 score (0.035 versus 0.054), and \flakylensplus exceeded it by less than its own fold-to-fold variability (0.070 versus 0.054). The weighted F1 score of neither variant exceeded the always-non-flaky baseline at the reported precision (0.958 and 0.948 versus 0.958). The per-fold flaky-class F1 score varied as much as its mean, so the small gap between the two variants does not show that one is better than the other.

On \flakebench, evaluated identically, the same variants recovered 261 of 263 flaky tests\footnote{The 280 flaky tests of \flakebench (\Cref{sec:datasets}) reduce to 263 under the $\leq$1024-token CodeBERT input filter we apply to match the \cidoft evaluation; the filtered set contains 263 flaky and \num{7985} non-flaky tests.} (flaky-class F1 scores of 0.984 and 0.967), so the test code carries a separable signal for that curated set. In \flakebench, most projects contribute exactly one flaky test (63\% of projects; a median of one per project), so its project-disjoint success rests on a signal shared across projects, whereas \cidoft concentrates flaky tests by construction (a median of 13 per project). A plausible reason is selection: \flakebench's flaky tests stem from \flakycat~\cite{akli2023flakycat}, which assigned a root cause category only where the cause could be recognized from the developer's fix and commit message, and the retained categories, such as unordered collections and async waits, are types whose flakiness tends to be apparent in the test code. The flakiness tracked in \idoft includes types that are not identifiable from the test code alone~\cite{DBLP:journals/corr/abs-2602-05465}. We therefore read the \flakebench result as lexical separability of a curated benchmark, not as evidence that source-only detection generalizes to developer-confirmed flakiness with rerun-confirmed non-flaky labels.
To complete the counterfactual square (the two labelings, fix-commit and rerun-confirmed, crossed with the two evaluation protocols), we additionally evaluated both variants on \cidoft under label-stratified random 10-fold cross-validation, the protocol class of the published evaluations.
Under this protocol, \flakifyplus reached a weighted F1 score of 0.989 with a flaky-class F1 score of 0.746 $\pm$ 0.030 (recall 0.641), and \flakylensplus 0.983 with 0.686 $\pm$ 0.109 (recall 0.709). \Cref{tab:rq2} reports these as the \cidoft{} (CV) rows. Both variants scored far above the constant baselines of the random folds (always-flaky flaky-class F1 score 0.047, always-non-flaky weighted F1 score 0.964) on the same data on which both variants had collapsed under project-disjoint evaluation: The protocol change alone moved the flaky-class F1 score from 0.035 to 0.746. Random cross-validation lets the model exploit within-project similarity, and this apparent detection ability does not transfer across projects.

To isolate the effect of the labeling alone, we also evaluated the fix-commit-labeled \idoft dataset (3,671 of 3,813 tests after the token filter) under the identical conventions (full test code, at most 1024 tokens, same fold generator and seed): \flakifyplus reached a weighted F1 score of 0.89 with a flaky-class F1 score of 0.937 $\pm$ 0.016, and \flakylensplus 0.83 with 0.878 $\pm$ 0.047. \flakifyplus's result lies in the published fix-commit range of 0.88--0.95, and on this flaky-majority dataset the always-flaky predictor alone attains a flaky-class F1 score of 0.916. We held the input conventions fixed rather than reproducing \flakify's test-smell preprocessing: The smell detector requires complete Java test-class sources, which are not part of \cidoft's model inputs, and it is Java-specific, whereas \cidoft also contains tests written in Kotlin and Groovy. Reproducing that preprocessing would add a dataset construction step with its own failure modes and, more importantly, weaken the experiment by reintroducing the empty method bodies it induces. Under project-disjoint folds, the same input stayed at its constant baseline: \flakifyplus{} reached a weighted F1 score of 0.79 and \flakylensplus{} 0.81 against an always-flaky fold mean of 0.78, with \flakifyplus{} predicting nearly every test flaky (flaky-class recall 0.996). We therefore compare the two labelings under one set of conventions. \Cref{tab:square} (\Cref{sec:discussion}) assembles the four cells and their baselines.

\begin{tcolorbox}[enhanced jigsaw,sharp corners, drop fuzzy shadow=ShadowColor]
\noindent\textbf{Answer RQ2}:
Our controlled counterfactual \cidoft confirms the findings of \textbf{RQ1}: On flaky tests drawn from the same benchmark, once the non-flaky class is rebuilt from reruns, both CodeBERT variants detected almost no flaky tests and did not meaningfully exceed any constant baseline (flaky-class F1 scores of 0.035 and 0.070 versus the always-flaky predictor's 0.054; the weighted F1 score not above the always-non-flaky baseline). Under the published cross-validation protocol, the same variants scored far above these baselines on the same data (flaky-class F1 scores of 0.746 and 0.686). Project separation removes this apparent detection ability. The same variants succeeded on the curated \flakebench (flaky-class F1 scores up to 0.984), so the test code carries a separable signal there but not for developer-confirmed flakiness with rerun-confirmed non-flaky labels.
\end{tcolorbox}

\begin{table}
\centering
  \caption{Results on \cidoft (the controlled counterfactual) and \flakebench (a curated benchmark). \(F1_f\) and Recall\(_f\) are for the flaky class; \(F1_w\) is the weighted F1 score. The constant rows (\Cref{sec:methodology}) are computed per held-out test fold from its flaky share; always non-flaky never predicts the flaky class, so its \(F1_f\) is undefined (--). All rows are project-disjoint except the two \cidoft{} (CV) rows, which report random cross-\allowbreak validation (CV): the (\cidoft, CV) cell of \Cref{tab:square}.}
  \label{tab:rq2}
  \small
  \setlength{\tabcolsep}{3pt}%
  \begin{tabular}{@{}llrrr@{}}
    \toprule
    \textbf{Dataset} & \textbf{Approach} & \textbf{\(F1_f\)} & \textbf{Recall\(_f\)} & \textbf{\(F1_w\)} \\
    \midrule
    \cidoft & \flakifyplus & 0.035 $\pm$ 0.058 & 0.021 & 0.958 \\
    \cidoft & \flakylensplus & 0.070 $\pm$ 0.078 & 0.120 & 0.948 \\
    \cidoft & Always flaky & 0.054 $\pm$ 0.020 & 1.000 & 0.002 \\
    \cidoft & Always non-flaky & -- & 0.000 & 0.958 \\
    \midrule
    \flakebench & \flakifyplus & 0.984 $\pm$ 0.028 & 0.992 & 0.999 \\
    \flakebench & \flakylensplus & 0.967 $\pm$ 0.029 & 0.992 & 0.998 \\
    \flakebench & Always flaky & 0.063 $\pm$ 0.009 & 1.000 & 0.002 \\
    \flakebench & Always non-flaky & -- & 0.000 & 0.951 \\
    \midrule
    \cidoft{} (CV) & \flakifyplus & 0.746 $\pm$ 0.030 & 0.641 & 0.989 \\
    \cidoft{} (CV) & \flakylensplus & 0.686 $\pm$ 0.109 & 0.709 & 0.983 \\
    \bottomrule
  \end{tabular}
\end{table}

\subsection{RQ3: Diagnosing Flaky E2E Failures (\flakeci)}
With the third research question, we aimed to determine the extent to which flaky failures in E2E tests require signals beyond the test source code. For this question, we first curated a dataset by mining CI logs from GitHub projects. We started by identifying repositories containing GitHub Actions workflow files that signal Playwright or Cypress test execution, yielding 522 Playwright-enabled and 377 Cypress-enabled repositories. We then collected CI run data for the 90 days preceding the observation date, the maximum period for which GitHub Actions retains execution logs. The collection windows fell between January and April 2026.
We labeled a test as flaky according to the criterion defined in \Cref{sec:methodology}, requiring explicit test-level evidence from the CI logs for both the failing and passing observations. The resulting dataset contains 86 flaky tests from 26 repositories: 74 Cypress tests from 22 repositories and 12 Playwright tests from 4 repositories.

\Cref{tab:rootcauses-rq3} summarizes the root cause categories assigned to the 86 flaky tests. For 42\% of them (36 of 86), we could attribute a cause from the test code and CI log. For the remaining 58\% (50 of 86), we could not attribute a cause without speculation, so these were recorded as \emph{Unknown}. Among the attributable cases, Network or external service instability was the largest category (27\% of all 86 tests). Two projects, \texttt{osmosis-frontend} and \texttt{infinispan-console}, account for 15 of the 23 Network cases.

Three cases show what the CI run captures and the test code does not. In \texttt{osmosis-frontend}, a swap test (\emph{User should be able to select stATOM/USDC}) failed with a 90-second timeout while waiting for token price data after opening the live application, failing 44 times and passing 3 on the same commit. In \texttt{bratislava.sk}, an E2E test of the site's official-board page failed because \texttt{cy.visit} returned HTTP 503 from the live site. In \texttt{processmaker/screen-builder}, a date-picker test failed on an invalid WebSocket URL \texttt{ws://localhost:undefined/} formed during application setup. In all three cases, the failure's cause was a runtime condition (a slow price feed, an unavailable site, an undefined port) that the test code does not show.

\begin{tcolorbox}[enhanced jigsaw,sharp corners, drop fuzzy shadow=ShadowColor]
\noindent\textbf{Answer RQ3}:
The test code and CI log sufficed to attribute a cause for 42\% of the flaky tests (36 of 86). Diagnosing the remaining 58\% would require execution evidence beyond these two artifacts. Network or external service instability was the largest attributable category (27\% of all 86 tests).
\end{tcolorbox}

\begin{table}
    \centering
    \caption{Root cause categories assigned to \flakeci's 86 flaky tests (74 Cypress + 12 Playwright from 22+4 repositories).}
    \label{tab:rootcauses-rq3}
    \begin{tabular}{@{}lrrrr@{}}
    \toprule
    \textbf{Cause category}
      & \textbf{Playwright}
      & \textbf{Cypress}
      & \multicolumn{2}{c}{\textbf{Both}} \\
    \cmidrule(l){4-5}
      & & & $\sum$ & \% \\
    \midrule
    Unknown & 3 & 47 & 50 & 58 \\
    Network / external service & 9 & 14 & 23 & 27 \\
    Environment / config & 0 & 7 & 7 & 8 \\
    Resource / state cleanup & 0 & 5 & 5 & 6 \\
    Async wait & 0 & 1 & 1 & 1 \\
    \midrule
    \textbf{Total}
      & \textbf{12}
      & \textbf{74}
      & \textbf{86}
      & \textbf{100} \\
    \bottomrule
    \end{tabular}
\end{table}

\section{Discussion}
\label{sec:discussion}

\textbf{On the replication results.} The results of our replication indicate that previous results on the \idoft benchmark are caused by the evaluation protocol and metrics used, the lack of proper baselines, and the simple syntactical patterns that sufficed to distinguish flaky from non-flaky tests in that data split, without any signal of flakiness itself. These patterns include preprocessing artifacts such as class-correlated empty test bodies. Additionally, we found the results of \flakify on \flakeflagger to be mostly explained by the data leakage problem that \citeauthor{DBLP:journals/pacmpl/Rahman0S25}~\cite{DBLP:journals/pacmpl/Rahman0S25} already described. Our reproduction after fixing the data leakage issue showed that the LLM performs only slightly better than our simple baseline. This finding is in line with prior work stating that the test code alone may not suffice as a signal to distinguish flaky from non-flaky tests in a meaningful scenario~\cite{DBLP:journals/corr/abs-2602-05465}.

\textbf{On the controlled counterfactual \cidoft.} The results on the controlled counterfactual, without shortcuts to distinguish flaky from non-flaky tests, support this claim.
Because \cidoft changes only the non-flaky labeling of the benchmark prior work used, this result cannot be attributed to a harder or unrelated dataset. We argue that our results align with the replication studies~\cite{DBLP:conf/iwpc/CamaraSEV21,DBLP:conf/msr/HabenHPCT21, DBLP:conf/esem/BerndtNB23} of the original lexical approach by \citeauthor{DBLP:conf/msr/0001MDdTB20}~\cite{DBLP:conf/msr/0001MDdTB20}, which have already shown that \codebased \classifier[s] lack generalizability when evaluated on real-world data. Our results suggest that these weaknesses cannot be mitigated by LLMs, as they are rooted in the task's framing. This holds beyond the CodeBERT-based detectors we evaluated: Prompting GPT-4o, GPT-OSS, and Qwen3-Coder classified flakiness from the test code only marginally better than random guessing~\cite{DBLP:journals/corr/abs-2602-05465}. As pointed out by prior work~\cite{DBLP:conf/icse/BerndtBB24}, when the number of repeated executions of a test goes toward infinity, every test can eventually yield a failing execution. For example, even central processing units (CPUs) have been shown to be susceptible to \enquote{silent data corruptions}, which propagate to the application level and may cause flaky failures~\cite{DBLP:journals/corr/abs-2102-11245}. Therefore, we argue that flakiness is not a static property that a test either has or does not have. It typically emerges from multiple contributing factors during runtime.

\textbf{On the analysis of \flakeci.} Our analysis of the newly curated \flakeci dataset showed that diagnosing most E2E flaky tests required execution evidence beyond the test code and CI logs. For most of the tests (58\%), the cause could not be attributed from the test code or the CI logs alone. This finding indicates that flakiness requires more in-depth debugging and observability tooling to pinpoint the root causes of flaky failures and inform decisions on whether a failure is caused by an actual defect or by flakiness.

\textbf{Implications for evaluating detectors.} Our reproduction points to a small set of reporting practices that would have exposed the fragilities we found. Detector studies should report the flaky-class precision, recall, and F1 score as primary metrics, together with their fold-to-fold variability, since a high weighted F1 score can coexist with near-zero flaky-class recall on imbalanced data. Such studies should evaluate their models under project-disjoint splits and against a majority-class baseline. A score that merely matches this baseline should not be read as detection. Input-aware trivial baselines, such as a linear model over token counts, and threshold-free metrics such as the area under the precision-recall curve~\cite{DBLP:conf/icml/DavisG06} would further tighten the attribution of high scores to labeling shortcuts rather than to detection ability. We leave both to future work. \Cref{tab:square} assembles the counterfactual square from \textbf{RQ1} and \textbf{RQ2}: both non-flaky labelings evaluated under each protocol, each cell against the strongest constant baseline of its dataset. We refer to a cell by its (labels, protocol) pair, such as (\idoft, CV). The two datasets differ in flaky share, so scores compare against their own baseline, not across rows. Under project-disjoint evaluation, neither labeling exceeds its baseline. Under random cross-validation, both do: Even with rerun-confirmed labels, the flaky-class F1 score rises from 0.035 in (\cidoft, Disjoint) to 0.746 in (\cidoft, CV), on identical data, once projects recur across folds. The cross-validation protocol thus produces scores above the constant baselines even without the labeling shortcut, while the shortcut remains visible within the protocol: In (\idoft, CV), the fix-commit labels reproduce the published range. Whatever signal the models exploit under cross-validation, with either labeling, does not transfer across projects. Non-flaky labels should come from repeated executions or CI history, with the observation protocol reported, rather than from fix-commit versions of flaky tests, which encode a near-duplicate shortcut. Preprocessing should likewise be checked for class-correlated artifacts such as empty test bodies before model scores are reported. Finally, a gain that appears only on a benchmark whose classes are lexically separable, such as \flakebench, should be reported as such, not as evidence of general flaky-test detection. Our replication package provides the project-disjoint pipelines, the \cidoft curation, and the \flakeci evidence bundles to support these practices.
\begin{table}
\centering
  \caption{Non-flaky counterfactual: weighted F1 scores of the \classifier[s] under the fix-commit (\idoft) and rerun-confirmed (\cidoft) non-flaky labelings and both protocols, random cross-validation (CV) and project-disjoint folds (Disjoint), together with each group's constant baseline and $\Delta$ (best value minus baseline). Rows name the \classifier and its input conventions: original (test-smell preprocessing, $\leq$512 tokens) or extended (full test code, $\leq$1024 tokens); \flakifyplus{} and \flakylensplus{} are our reimplementations (\Cref{sec:methodology}). Original-row values are our reproductions, with $^{*}$ marking the data-leakage fix and the published score in parentheses.}
  \label{tab:square}
  \small
  \setlength{\tabcolsep}{3pt}
  \begin{tabular}{@{}lll@{}}
    \toprule
     & \textbf{CV} & \textbf{Disjoint} \\
    \midrule
    \textbf{Fix-commit (\idoft)} & & \\
    \enspace \flakify{} (original) & 0.95\,(0.94),\,0.88$^{*}$ & 0.80$^{*}$ \\
    \enspace \flakyq{} (original) & 0.95\,(0.93),\,0.88$^{*}$ & 0.80$^{*}$ \\
    \enspace \flakifyplus{} (extended) & 0.89 & 0.79 \\
    \enspace \flakylensplus{} (extended) & 0.83 & 0.81 \\
    \enspace \emph{baseline (always flaky)} & \emph{0.78} & \emph{0.80} \\
    \enspace $\Delta$ (best $-$ baseline) & \textbf{+0.18} & \textbf{+0.01} \\
    \addlinespace
    \textbf{Rerun-confirmed (\cidoft)} & & \\
    \enspace \flakifyplus{} (extended) & 0.989 & 0.958 \\
    \enspace \flakylensplus{} (extended) & 0.983 & 0.948 \\
    \enspace \emph{baseline (always non-flaky)} & \emph{0.964} & \emph{0.958} \\
    \enspace $\Delta$ (best $-$ baseline) & \textbf{+0.025} & \textbf{+0.000} \\
    \bottomrule
  \end{tabular}
\end{table}

\textbf{Future work.} Our results reframe flaky-test prediction as a question about executions rather than test code, and the instruments we release make that question actionable. We see three directions for the community, each grounded in a finding or an artifact of this study.

\textbf{Failure-level classification.} Instead of labeling a test, the field should move to classifying whether an individual observed \emph{failure} is flaky from its execution evidence, such as logs, stack traces, and CI context. \flakeci offers an initial labeled corpus, and because our same-commit mining reads existing CI logs rather than re-running suites, it reaches the slow, complex systems that rerun-based labeling cannot, countering rerun-based labeling's bias toward projects that are easy to build and re-execute. The open questions are which runtime features separate flaky from genuine failures and whether they transfer across projects, languages, and frameworks.

\textbf{Environment-conditioned failure probability.} The binary test-level label should be replaced by a calibrated probability that a test fails given its execution environment, including the runner, operating system, browser, and network conditions. Such an estimate turns rerun budgeting into a decision under uncertainty, where how often to re-execute a failing test follows from its estimated failure probability and the asymmetric costs of a blocked pipeline against a shipped defect. This ties prediction to the compute cost that motivates flakiness handling, and measuring how much failure-level triage lowers that cost in industrial pipelines is itself an open question.

\textbf{Language models that drive interventions.} Our results discourage static LLM \classification, but not the use of language models themselves. The evidence we find decisive, namely logs, stack traces, and CI context, is exactly the unstructured artifact such models read well. We see a promising direction in agents that move from reading to experimenting. Given a failing run, such an agent hypothesizes a cause and then perturbs the execution environment to test it, for example by injecting network latency or loss, adding CPU or memory pressure, shifting the system clock, or reordering tests, while observing which perturbations change the outcome. \citeauthor{DBLP:conf/icse/TerragniSF20} explored such non-deterministic dimensions by re-executing flaky tests in dedicated containers~\cite{DBLP:conf/icse/TerragniSF20}; the agent would select the perturbations from the failure evidence rather than enumerating them. This places the model inside the execution loop that static \classification lacked, using it to characterize a test's failure distribution rather than to guess a label from source code.

\section{Threats to Validity}
\label{sec:threats}
We address threats to our study's internal, external, conclusion, and construct validity.
\subsection{Internal Validity}
We describe the degree to which we can rule out alternative explanations for our results~\cite{brewer2014research}.

\textbf{On the reproduction of existing studies.} We relied on the author-supplied replication packages when reproducing the existing pipelines of \flakify, \flakyq, and \flakyxbert. Executing them locally required minor source code adjustments, such as updating paths and dependencies. Furthermore, unlike the authors of the studies, we did not run the experiments on CUDA devices but on a Mac Studio with Metal Performance Shaders (\texttt{MPS}), which might explain slight variations in the results of our experiments. Two observations limit the impact of this threat: The environment-induced variation stayed within 0.02 in our reproductions, an order of magnitude smaller than the effects our conclusions rest on, and the qualitative analysis of the underlying datasets ties the large drops after the leakage fix to dataset properties rather than to the execution environment.

\textbf{On repeated executions for non-flaky labels.} We conducted all repeated executions on a single CI configuration. Thus, flakiness triggered only under different OS versions, locale settings, or resource pressure~\cite{DBLP:journals/tse/SilvaGGATdLWB24} would be missed, potentially leading to tests being mislabeled as non-flaky. This limitation is inherent to rerun-based labeling rather than specific to our setup, as even \flakeflagger's \num{10000} reruns showed (\Cref{sec:datasets}): Rarely failing tests lead to underestimating the number of flaky tests~\cite{DBLP:conf/icse/GruberRPSM024}, and the executions needed to expose low-probability failures have been quantified~\cite{DBLP:journals/corr/abs-2601-08998}. For a test with a per-run failure probability of 1\%, our 500-run budget observes at least one failure with a probability of 99.3\%. Rarer failures are more likely to go unobserved and be mislabeled as non-flaky. A model that correctly flags such a test is then charged with a false positive, which lowers the measured flaky-class precision and F1 score. The flaky and non-flaky classes also carry labels from different oracles: developer confirmation yields high-precision flaky labels, and repeated execution is the established method for confirming non-flaky labels. The near-zero flaky-class recall is computed over the developer-confirmed flaky tests, so mislabeled non-flaky tests do not enter that measurement, and our budget exceeds the 100 reruns common in prior curation~\cite{DBLP:conf/msr/0001MDdTB20,DBLP:journals/pacmpl/Rahman0S25}.

\textbf{On the \cidoft construction.} Relative to \idoft, \cidoft narrows the flaky class to the developer-confirmed tests of the 57 buildable projects (\num{1319} of \num{3195} tests) and shifts the flaky share from 84\% to 2.4\%. The collapse we report on \cidoft therefore reflects the labeling change together with these population shifts and the protocol change, which the counterfactual square separates (\Cref{tab:square}). On identical data and conventions, the protocol change alone moves the flaky-class F1 score from 0.746 in (\cidoft, CV) to 0.035 in (\cidoft, Disjoint), so we attribute the published scores to the labeling shortcut and the cross-validation protocol jointly rather than to the labeling alone. The \(\leq\)1024-token input filter (\Cref{sec:methodology}) removes only \num{1536} of the \num{56004} extracted tests (2.7\%); the \textbf{RQ2} comparison applies the same filter to \flakebench, so the performance difference between the two datasets cannot stem from the filter. The retained flaky tests, however, are the developer-confirmed subset, the highest-precision positive labels in \idoft, which should make detection easier rather than harder. The 2.4\% flaky share matches the 2--4\% prevalence of the rerun-labeled datasets; a detector deployed on a full test suite would face a similarly small flaky minority.

\textbf{On the manual root cause labeling.} A single author categorized the 86 flaky tests in \flakeci, which risks subjective assignments. We mitigated this by defining the category set and procedure with a second author and by anchoring every assignment to the matched failing and passing log evidence recorded in our replication package. When the evidence did not identify a cause, we assigned \emph{Unknown} rather than speculate, so our reported 42\% is a lower bound. We do not claim the remaining failures are undiagnosable from the code and logs, only that our manual analysis did not find a cause there. A same-commit pass/fail pair could still reflect deliberately different configurations rather than non-determinism, which our same-workflow-and-job criterion reduces (\Cref{sec:methodology}). For \flakeci's purpose here, determining whether failures can be diagnosed from the test code and CI logs, a single expert coder over 86 tests is adequate; a benchmark for training would require multiple independent coders and broader coverage.

\subsection{External Validity}
We describe the generalizability of our results~\cite{DBLP:journals/ese/BaltesR22}.

\textbf{On the benchmarks for \codebased \classification.} The datasets used to answer \textbf{RQ1} and \textbf{RQ2} focus on Java projects using the Maven or Gradle ecosystem. Flakiness patterns in other ecosystems may differ in distribution and root causes, which would limit cross-language generalization. The detector results are therefore established only for unit tests in JVM languages, nearly all of them Java. \flakeci covers JavaScript/TypeScript E2E tests, but we use it only for the diagnosis question of \textbf{RQ3}, not to evaluate the \classifier[s]. \cidoft contains only projects we could build and re-execute 500 times, so the \textbf{RQ2} findings come from projects that are comparatively easy to build and test. This selection works in the \classifier[s]' favor: The collapse appears on exactly the projects that are easiest to build, execute, and analyze. The restriction to Java projects with Maven or Gradle matches the setting in which the reproduced \classifier[s] and their benchmarks were developed, so it does not disadvantage them either.

\textbf{On the curation of \flakeci.} We curated the \flakeci dataset from a 90-day CI mining window, the maximum log retention of GitHub Actions. Thus, the dataset may capture seasonal patterns of that period rather than general problems. Furthermore, we focused on GitHub Actions runners, which might not generalize to other CI environments. The corpus is also small and concentrated: 86 tests from 26 repositories, 74 of them Cypress, and 15 of the 23 Network cases stem from two projects, so the reported category shares are indicative rather than representative. For its purpose here, probing whether causes can be attributed from the test code and CI logs, verified same-commit pass/fail evidence matters more than breadth; each of the 86 tests has both a failing and a passing run on the same commit.

\subsection{Conclusion Validity}
We describe the degree to which our data support the conclusions we draw~\cite{wohlin2024experimentation}.

Where our design yields paired observations, we tested the direction of each comparison with an exact two-sided sign test~\cite{wohlin2024experimentation}. Fixing the data leakage lowered the weighted F1 score in 9 of the 10 cross-validation folds for both detectors ($p$ = 0.02 each), and on \flakeflagger the always-flaky baseline exceeded the classifier's per-project flaky-class F1 score in 19 of 23 projects ($p$ = 0.003). The four folds of \textbf{RQ2} are too few for such a test. At the reported precision, two comparisons are equal (0.80 versus 0.80 and 0.958 versus 0.958). The unrounded differences, 0.004 for the \textbf{RQ1} Disjoint classifier over the held-out baseline and 0.0001 for \flakifyplus's weighted F1 score over the always-non-flaky baseline, are far below the fold-to-fold variability of 0.02--0.08. The \textbf{RQ3} percentages are based on only 86 tests from 26 repositories and are correspondingly uncertain.

\subsection{Construct Validity}
We describe the degree to which our scales and metrics measure the intended properties~\cite{DBLP:conf/ease/RalphT18}.

\textbf{On project-disjoint evaluation as a proxy for practical use.}
Our evaluation protocol holds out a set of projects as the test set on which the models are never trained. This simulates cross-project use but may not capture within-project temporal drift, which could be closer to practical use. However, we stress that our evaluation protocol aims to show that models struggle to learn more abstract features that underlie flakiness, instead relying on lexical features, as \citeauthor{DBLP:conf/iwpc/CamaraSEV21}~\cite{DBLP:conf/iwpc/CamaraSEV21} already showed for the original lexical approach.

\textbf{On the scope of \codebased \classifier[s].} Our claim targets detection from the test source code. The classifiers we evaluated see only test-method tokens. One might argue that production code called by the test, configuration files, or dependency manifests are also ``code''. Our construct draws the boundary at the test method, matching prior work but potentially excluding useful static signals. We argue, however, that our analysis of \flakeci supports this boundary: While a potential dependency on an external service, DOM state, or setup step may be statically visible, deciding whether it explains a particular flaky failure requires execution-level evidence.

\section{Related Work}
\label{sec:related}

\textbf{Benchmark and evaluation-validity critiques.} Our work continues a line of recent studies questioning whether reported gains on \codebased flakiness benchmarks reflect real \classification ability. Analogous benchmark shortcuts are documented for other SE-ML tasks, such as deep-learning vulnerability detection~\cite{DBLP:journals/tse/ChakrabortyKDR22}.
\citeauthor{DBLP:journals/pacmpl/Rahman0S25} identified the cross-validation data-leakage flaw described in \Cref{sec:background}~\cite{DBLP:journals/pacmpl/Rahman0S25}; correcting it and re-evaluating on a more realistic dataset (\flakebench) with rerun-confirmed non-flaky labels lowered the measured performance.
We found that the same flaw affected the leave-one-project-out evaluation.
\citeauthor{DBLP:journals/corr/abs-2602-05465} argue more directly that the test code may not contain the information needed to distinguish flaky from non-flaky tests, finding that whether humans consider themselves able to label a test from its code depends on the flakiness type, with classes such as side effects of tested or utility code remaining invisible in the test itself~\cite{DBLP:journals/corr/abs-2602-05465}.
Our contribution differs from both.
\citeauthor{DBLP:journals/pacmpl/Rahman0S25} localized one implementation defect, and \citeauthor{DBLP:journals/corr/abs-2602-05465} showed through human labeling that the test code \emph{may} lack the signal for certain flakiness classes.
We instead show that the collapse to the baseline is not specific to one \classifier or one labeling: It recurs across the two published \codebased \classifier[s] that permit project-disjoint evaluation and persists on \cidoft, a controlled counterfactual that removes the labeling shortcut while keeping the benchmark's flaky tests.
We then move beyond showing that the code lacks the signal to locating it, finding that most flaky E2E failures cannot be diagnosed from the test source and the CI log alone and that, where they can, the decisive evidence sits in the CI run rather than in the test source.

\textbf{Detection from signals other than test code.} Much work detects flakiness without inspecting test code, which supports our thesis that the relevant signal is elsewhere.
Several approaches model test execution history. \citeauthor{DBLP:conf/icse/KowalczykNGSLM20} ranked tests at Apple using entropy and flip-rate computed over pass/fail histories~\cite{DBLP:conf/icse/KowalczykNGSLM20}, and \citeauthor{DBLP:conf/icse/HerzigN15} mined association rules over test-step results to flag false alarms~\cite{DBLP:conf/icse/HerzigN15}. \citeauthor{DBLP:conf/kbse/HabenHMHPCT24} argue that prediction should account for execution failures and target flaky \emph{executions} rather than flaky tests~\cite{DBLP:conf/kbse/HabenHMHPCT24}.
Dynamic, rerun-based detectors such as \idflakies{} identify flaky tests by reordering and re-executing test suites~\cite{LamETAL19iDFlakies}, and log-based techniques locate the root cause of flakiness by diffing the logs of passing and failing runs in industrial settings~\cite{LamGNST19}.
Surveyed developers nonetheless still asked for lightweight static \classification support~\cite{DBLP:conf/icst/GruberF22}.

\textbf{Empirical studies of flakiness causes.} Root-cause studies attribute most flakiness to environmental rather than syntactic causes.
\citeauthor{DBLP:conf/sigsoft/EckPCB19} had Mozilla developers classify previously fixed flaky tests and reported a taxonomy led by concurrency and async-wait issues, alongside newly identified categories such as overly restrictive value ranges~\cite{DBLP:conf/sigsoft/EckPCB19}. \citeauthor{DBLP:conf/icse/RomanoSGYW21} found that UI-based (E2E) flaky tests are dominated by async waits and environment issues such as network, resource, and platform behavior~\cite{DBLP:conf/icse/RomanoSGYW21}.
These are causes a \codebased \classifier largely cannot observe in the expensive E2E setting that \textbf{RQ3} targets.
Such flakiness is costly in practice: An industrial CI case study attributed at least 2.5\% of productive developer time to handling flaky failures~\cite{DBLP:conf/icst/LeinenEPSSJ24}.

\textbf{Toward failure-level triage.} Our forward-looking position, moving from labeling tests to classifying observed failures, aligns with work that operates at the execution level.
\citeauthor{AlshammariAHB24} classified whether an individual test \emph{failure} is flaky from its failure output, such as stack traces, rather than labeling the test from its source~\cite{AlshammariAHB24}, complementing the log-based root cause analysis of \citeauthor{LamGNST19}~\cite{LamGNST19} and the alarm-level detection of \citeauthor{DBLP:conf/icse/HerzigN15}~\cite{DBLP:conf/icse/HerzigN15}. Researchers at Google have trained a machine learning model to detect flaky failures and suppress re-execution of tests~\cite{DBLP:conf/ftw/HoangB24}, and \citeauthor{DBLP:conf/icsm/AnY0HY24} used failure symptoms, such as error messages and stack traces, to identify flaky failures in SAP HANA's CI pipeline~\cite{DBLP:conf/icsm/AnY0HY24}. \citeauthor{DBLP:conf/sigsoft/LampelJAZ21} classified intermittent job failures from CI telemetry at Mozilla~\cite{DBLP:conf/sigsoft/LampelJAZ21}, and \citeauthor{DBLP:conf/icse-seip/AidassoBT25} prioritized flaky job-failure categories at {TELUS}~\cite{DBLP:conf/icse-seip/AidassoBT25}. \citeauthor{DBLP:conf/kbse/HabenHMHPCT24}'s study cited above, a collaboration with Ubisoft, Meta, and Broadcom, makes this case at industrial scale~\cite{DBLP:conf/kbse/HabenHMHPCT24}.
We view this execution-grounded framing, rather than \codebased test-level \classification, as the better-posed problem.

\section{Conclusion}
\label{sec:conclusion}
We conducted a study of \codebased flaky test \classification by reproducing three published LLM-based \classifier[s] (\flakify, \flakyq, \flakyxbert) and evaluating them under progressively stricter protocols. Our reproduction revealed that the reported F1 scores of these \classifier[s] rest on evaluation shortcuts in the commonly used benchmarks, including data leakage across cross-validation folds, near-duplicate flaky/non-flaky test pairs introduced by the fix-commit labeling strategy, and project overlap between training and test splits. After we addressed these issues and evaluated in a project-disjoint setting, \flakify and \flakyq collapsed to the level of a majority-class baseline, while rerunning \flakyxbert's four affected notebooks on their own data lowered its weighted F1 score from 0.94 to 0.83, which stays above the always-flaky baseline of 0.66 in the within-project setting. We conclude that the high F1 scores reported in prior work reflect properties of the benchmarks rather than a learned ability to distinguish flaky from non-flaky tests.

To investigate this idea, we curated \cidoft, a controlled-counterfactual dataset of \num{54468} tests from 57~projects that rebuilds the non-flaky class of the benchmark prior \classifier[s] were evaluated on, deriving non-flaky labels from 500 repeated executions rather than fixed versions of flaky tests. Evaluating both CodeBERT variants on \cidoft in a project-disjoint setting confirmed the findings of our reproduction: Neither meaningfully outperformed the constant baselines. These results are consistent with prior replication studies of lexical \codebased approaches without LLMs, which already showed limited generalizability on unseen projects~\cite{DBLP:conf/iwpc/CamaraSEV21}, on newer revisions of the same projects~\cite{DBLP:conf/msr/HabenHPCT21}, and on industrial data~\cite{DBLP:conf/esem/BerndtNB23}. Our results suggest that replacing lexical features and traditional machine learning with LLMs does not overcome this limitation, as it is rooted in the task framing rather than in model capacity. Prompting GPT-4o-class LLMs classified flakiness from the test code only marginally better than random guessing~\cite{DBLP:journals/corr/abs-2602-05465}.

We complemented this quantitative analysis with a root cause study of flaky E2E tests we curated from CI logs on GitHub Actions. Our manual classification of 86~flaky tests showed that for 58\% of the tests, we could not attribute the failure to a cause from the test code or the CI log. Therefore, we conclude that the test code records what is asserted, not whether the runtime conditions for that assertion will hold on a given execution, which helps explain why \codebased \classifier[s] struggle to reliably detect flakiness.

Our results relocate flaky-test prediction. While predicting flakiness from static code alone is appealing for its low cost, our results show that, for many kinds of flakiness, the test code does not contain the information this prediction requires. Instead of asking whether a \emph{test} is flaky, we can ask whether an observed \emph{failure} is flaky, or how likely a test is to fail given its execution environment, questions grounded in the execution-level evidence that our \flakeci analysis found decisive in every diagnosed case. Our two datasets support work on these questions. \cidoft is a control on which static \classifier[s] no longer achieve inflated scores, and \flakeci is a corpus of confirmed flaky E2E failures with the diagnostic evidence attached. We encourage the community to extend such corpora into failure-level benchmarks by adding genuine-failure counterexamples, which the same mining method can collect. We lay out concrete openings for this agenda in \Cref{sec:discussion}, including failure-level classification and language-model agents that probe the execution environment directly. A nascent line of work already argues for predicting flaky \emph{executions} rather than flaky \emph{tests}~\cite{DBLP:conf/kbse/HabenHMHPCT24,AlshammariAHB24}.
Our findings give that direction empirical support and new shared resources.
How far are we from detecting flaky tests? Farther than reported F1 scores suggest. Under project-disjoint evaluation of developer-confirmed flakiness with rerun-confirmed non-flaky labels, \codebased \classifier[s] are indistinguishable from a majority-class baseline, and for most flaky E2E tests we examined, the test code and CI log did not suffice to attribute the cause. The way forward, we argue, is to reframe the question at the level of executions.

\bibliographystyle{IEEEtranN}
\bibliography{references}

\end{document}